\documentclass[conference]{IEEEtran}
\IEEEoverridecommandlockouts
\usepackage[english]{babel}
\usepackage{blindtext}
\usepackage{lipsum}
\usepackage{cite}
\usepackage{algorithmic}
\usepackage{graphicx}
\usepackage{xcolor}
\usepackage{etoolbox}

\makeatletter
\def\temp{dvips.def}
\ifx\Gin@driver\temp
\def\Ginclude@graphics#1{\def\temp{#1}---image \expandafter\strip@prefix\meaning\temp---}
\fi
\makeatother

\usepackage[cmex10]{amsmath}

\DeclareMathOperator{\E}{E}

\usepackage{array}

\usepackage{url}

\usepackage{xspace}

\newcommand{\ie}{\emph{i.e.}\xspace}

\newcommand{\figref}[1]{Fig.~\ref{#1}}
\newcommand{\secref}[1]{\ref{#1}}

\newcommand{\tabref}[1]{Table~\ref{#1}}

\begin{document}
%

\title{Stateless multicast switching in software defined networks}

\author{\IEEEauthorblockN{Martin J. Reed, Mays Al-Naday, Nikolaos Thomos}%
\IEEEauthorblockA{
University of Essex\\
Colchester, CO4 3SQ, UK\\
Email:\{mjreed,mfhaln,nthomos\}\\
@essex.ac.uk}%
\and%
\IEEEauthorblockN{Dirk Trossen}%
\IEEEauthorblockA{InterDigital Europe, Ltd.\\
London, EC2A 3QR, UK\\
Email: Dirk.Trossen\\
@InterDigital.com}%
\and
\IEEEauthorblockN{George Petropoulos, Spiros Spirou}%
\IEEEauthorblockA{Intracom SA Telecom Solutions, \\
Peania, 19002, Greece\\
Email:\{geopet,spis\}\\
@intracom-telecom.com}%
\thanks{This work was carried out within the project POINT, which has received funding from the European Union's Horizon 2020 research and innovation programme under grant agreement No 643990}}


%


\maketitle

\begin{abstract}
Multicast data delivery can significantly reduce traffic in operators' networks, but has been limited in deployment due to concerns such as the scalability of state management. This paper shows how multicast can be implemented in contemporary software defined networking (SDN) switches, with less state than existing unicast switching strategies, by utilising a Bloom Filter (BF) based switching technique. Furthermore, the proposed mechanism uses only proactive rule insertion, and thus, is not limited by congestion or delay incurred by reactive controller-aided rule insertion. We compare our solution against common switching mechanisms such as layer-2 switching and MPLS in realistic network topologies by modelling the TCAM state sizes in SDN switches.  The results demonstrate that our approach has significantly smaller state size compared to existing mechanisms and thus is a multicast switching solution for next generation networks.  \end{abstract}

\section{Introduction}

Multicast routing and switching is generally more complex than its unicast counterpart, predominantly because: optimal multicast routing is an NP-complete problem (a Steiner tree optimisation)~\cite{Kompella1993}; and, IP routers/switches have to maintain specific multicast state~\cite{Martinez-Yelmo2007}. Although multicast is not, truly, deployed as an inter-domain service at Internet scale~\cite{Careglio2014}, it is widely deployed at intra-domain level for applications such as IPTV. While multicast has problems at the IP layer, these difficulties are somewhat greater at lower layers that do not natively support multicast, such as multi-protocol label switching (MPLS). RFC6513~\cite{Rosen12} has added multicast support to MPLS; however, at operator scale, a trade-off between optimality of routing and scalability of state is needed, as described by Martinez-Yelmo \emph{et al.} ~\cite{Martinez-Yelmo2007}. 

The problems with deploying multicast at operator scale has motivated stateless solutions, such as multi-protocol stateless switching (MPSS)~\cite{Zahemszky2010} and line speed publish/subscribe inter-networking (LIPSIN), that use the Bloom filter (BF) as the forwarding identifier (FID) in the packet-header. For brevity, this form of switching will be described as \emph{BF switching}. Indeed BF switching has been utilised by architectures such as PSIRP and PURSUIT~\cite{Fotiou10,Trossen2012} that form the basis of a clean-slate information centric network (ICN)~\cite{Xylomenos2014}. These efforts demonstrate that BF switching can deliver very efficient unicast/multicast forwarding with minimal state, however, they require either software-based switching or custom hardware.

This paper follows a BF approach, implemented natively in SDN, as a new technique for providing stateless multicast switching at operator scale. While a BF has been previously proposed for stateless multicast switching~\cite{Jokela09,Zahemszky2010}, this paper is the first to show that it can be implemented directly in contemporary SDN switches, without reactive controller intervention. We focus on an SDN solution as it is a frequent proposal for next generation networks at operator scale~\cite{Kreutz2015}. More specifically, SDN is proposed as a ``clean-slate'' network upgrade, replacing technologies such as spanning-tree or OSPF in data-centre or operator networks respectively. The term ``software defined networks'' can encompass a number of different software controlled network technologies; however, in this paper the term SDN will be used, specifically, to describe contemporary layer-2/3 Ethernet switches controlled by a centralised controller employing the OpenFlow protocol~\cite{OpenFlow1.2}.

There have been previous attempts at using SDN for clean-slate proposals such as ICN. Chanda and Westphal proposed \emph{ContentFlow}, which maps application layer content information onto a legacy IP transport mechanism using extensions to an SDN controller~\cite{Chanda2013}. Alternatively, Syrivelis \emph{et al.} described in~\cite{Syrivelis2012} how SDN can be used for forwarding in the PURSUIT architecture by using the BF switching header within the SDN controller to route packets to the appropriate destination on a packet-by-packet basis. Notably, neither of these propositions have provided scalable solutions at operator network sizes, as the controller has to react to each application session or packet, respectively. In contrast, the proposition here directly implements the BF switching in the SDN switches. 

The rest of this paper is structured as follows: Section~\ref{sec:proposition} will describe the proposal together, with Section~\ref{sec:algo} providing an analytical approach to the bounds of the SDN flow sizes; Section~\ref{sec:results} demonstrates the scalability advantages of the solution by modelling and comparing the flow entry requirements in realistic network topologies; finally, Section~\ref{sec:conclusion} briefly concludes and outlines future work.

\section{Overview of proposition}
\label{sec:proposition}

\subsection{Generalised description of BF switching}
\label{sec:bf-switching}

\begin{figure}[tb]
  \centering
  \includegraphics[width=\linewidth]{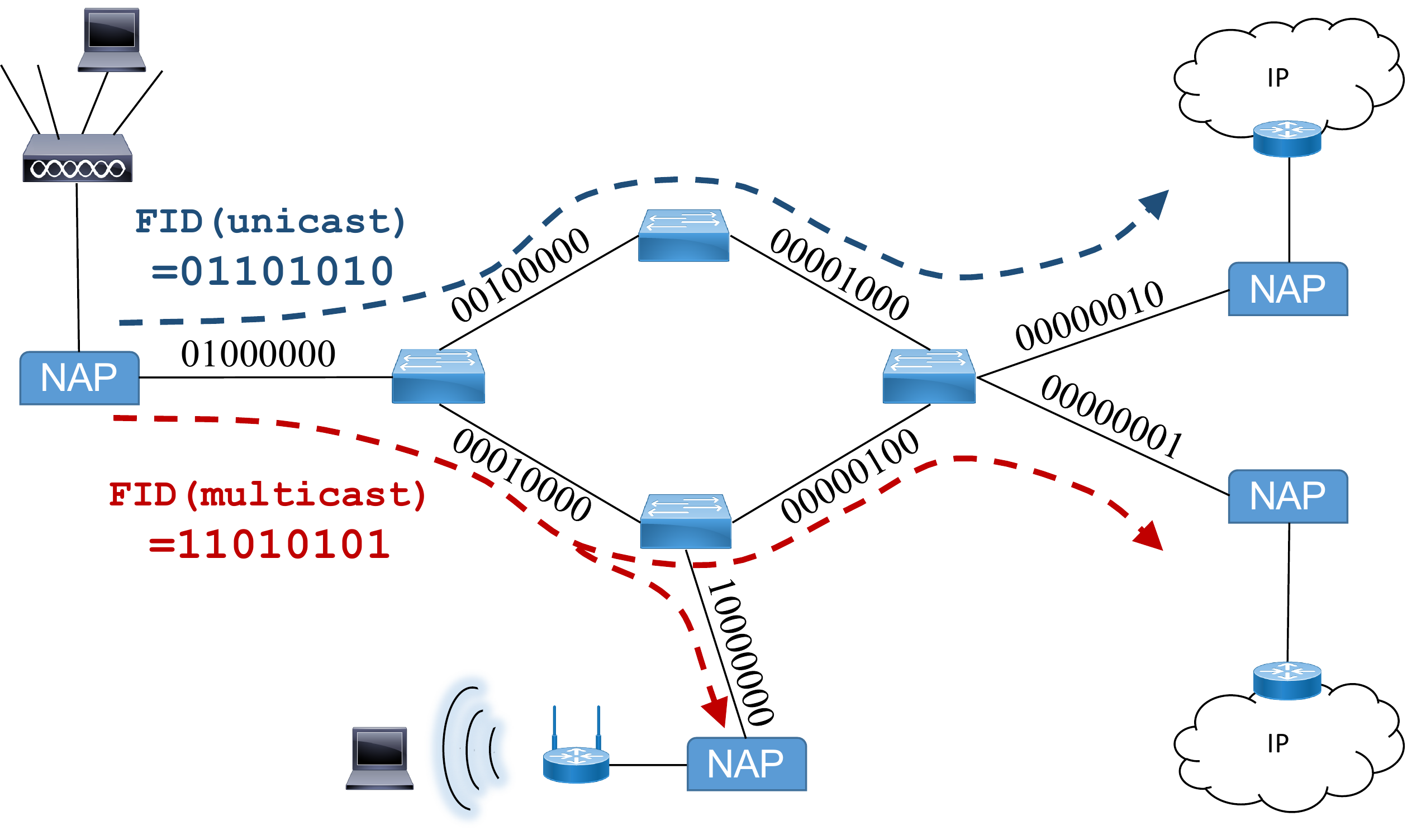}
  \caption{Highly simplified network architecture showing an example of BF switching}
  \label{fig:arch}
\end{figure}

A highly simplified form of the target, intra-domain, operator network architecture is illustrated in~\figref{fig:arch}. Network attachment points (NAPs) are shown interfacing the internal network, that uses BF switching, to external networks that use conventional IP routing. The architecture in~\figref{fig:arch} is similar to that proposed by MPSS~\cite{Zahemszky2010}, or by the EU project POINT~\cite{Trossen2015}. With MPSS the NAP would be integrated with the provider edge (PE), mapping IGMP onto BF switching. Alternatively, POINT integrates the NAP into any suitable edge router and performs IP-to-ICN convergence. At the edges, IP devices use the core network without knowledge of the SDN/BF core switching. 

In its simplest form, BF switching operates by assigning fixed length link identifiers (LIDs) for each link in the network, as described by~\cite{Jokela09}.  In \figref{fig:arch}, each link is assigned an 8-bit LID. In practice, to avoid false positives, the LID is much larger, typically in the order of hundreds of bits. The mechanism merges the LIDs on a path/tree to form a BF of the same length as the LIDs; this becomes the FID used to forward the packet.  Formally, we describe BF switching by considering a network represented as a graph $G(V,E)$ with an LID, $L(e)$, associated with each edge $e\in E$. The FID $F_P$ of a path (or tree) with $n$ edges, $P=\langle e_1, \ldots, e_n | e \in E \rangle$, is given by: 
\begin{equation} 
\label{eq:fid}
F_P = L(e_1) \vee L(e_2) \vee \ldots \vee L(e_n)
\end{equation}
At a node $v \in V$ of degree $d_v$ with outgoing edges $\Omega_v=\{e_1,\ldots,e_d\}$, we may represent the forwarding decision $D_e$ for an outgoing edge $e$ with LID $L(e)$ as: \begin{equation}
  \label{eq:bf-match}
  D_e= \left\{
  \begin{array}{ll}
    \text{true}, & \text{if } L(e) = L(e) \wedge F_p \\
    \text{false}, & \text{otherwise}
  \end{array} 
\right.
\end{equation}
\figref{fig:arch} shows examples of a unicast path and a multicast tree (with two destinations), and the associated FIDs created using~\eqref{eq:fid}. These examples highlight an inherent advantage of BF-encoding of network paths, namely the instantaneous formation of multicast relations based on individual FIDs. In other words, the knowledge of individual unicast FIDs from a source to a number of destinations allows for creating temporary multicast FIDs by simply ORing the individual unicast FIDs. The resulting FID encodes the multicast tree from the source to the selected destinations, with the OR operation being performed locally at any element that knows of the individual FIDs, such as the NAPs in \figref{fig:arch}. In POINT~\cite{Trossen2015} this is exploited to offer a capability called \emph{coincidental multicast}, in which responses to individual quasi-synchronous HTTP requests are delivered to the clients using multicast rather than individual unicast responses, resulting in potentially significant improvements of network utilization.

Note that, the LIDs need to be chosen carefully to avoid false-positives~\cite{Jokela09}, which result in traffic being sent over links that were not defined in the path or multicast tree. If the number of links $|E|$ is smaller than the bit-width of the LIDs, then setting one mutually exclusive bit in each LID is sufficient to guarantee no false positives. As noted in ~\cite{Zahemszky2010}, the use of constant length BF identifiers, however, limits the scalability of the network graph that can be encoded. To remove this limitation, we foresee a solution that divides an autonomous forwarding system into pre-defined zones, with a fixed length FID used per-zone. The FID in the header matching field is used to forward from a zone's ingress to its egress, while including all other zone FIDs in an FID-\emph{list} (FLIST) within the packet as it traverses the zone. The switches located at zone boundaries copy the appropriate FID from the FLIST into the matching field that is used in the forwarding operation of the next zone. With such a mechanism, any network size can be accommodated in an extensible manner, while utilizing the basic forwarding operation presented in this paper. In the remainder of this paper, we will assume a single such zone, for the sake of simplicity, and leave the details of such an extensible solution for future work.

\subsection{SDN switching}
\label{sec:sdn-switching}

Ethernet switching has evolved in two areas of interest to this work: wider, more complex, content addressable memory (CAM); and the separation of data and control planes~\cite{Kreutz2015}.

\emph{CAM evolution} has meant that binary CAM, originally used for exact matching as required for Ethernet destination MAC switching, has been widened in length to incorporate further fields in the headers. A typical implementation now encompasses MAC source \& destination, VLAN tag, MPLS label, IP(v4 or v6) source \& destination and TCP/UDP source \& destination ports. To enable wildcard matching, as required by IP longest prefix match, or port ranges in access control lists (ACLs), certain fields in the header use ternary CAM (TCAM). It is important to note that the TCAM is one of the most expensive parts of an ASIC switch implementation in terms of silicon and power requirements.

\emph{Separation of control and data planes} has allowed switching decisions to be made according to operator requirements, instead of restrictions derived from in-band control protocols. One of the main enabling technologies for this, is the development of a centralised controller that uses the OpenFlow protocol~\cite{OpenFlow1.2} to control the TCAM entries in the switch and the associated actions for packets that match the flow entries in the TCAM. A good example of the optimisations that this permits is shown in the L2switch implementation of the OpenDaylight controller\cite{OpenDaylight}. Instead of using spanning-tree to limit traffic to a loop-free tree, L2switch finds shortest-paths and inserts flow entries for Ethernet source/destination pairs in switches along the paths. This allows the operator to, potentially, utilise all links in the network. This L2switch implementation will be used as one of the evaluation comparisons in this paper. It should be noted that the L2switch requires that the first packet in any new communication flow between a source/destination pair is sent to the controller, which incurs a delay~\cite{Kreutz2015}.

\subsection{Implementing BF switching with SDN}
\label{sec:bf-sdn}

To understand the method for directly implementing BF switching in SDN, consider the use of the TCAM for matching a field of value $f$ that represents a range of values using a mask $m$. For example, the IPv4 longest prefix match $f/m$=\texttt{192.168.8.0/21}, where \texttt{/21} represents a mask $m$=\texttt{255.255.248.0}, would match all IP addresses \texttt{192.168.8.0} -- \texttt{192.168.15.255}. The TCAM implements a decision operation $D$ on the appropriate packet header $h$:
\begin{equation}
  \label{eq:ip4-match}
  D(h,f,m)= \left\{
  \begin{array}{ll}
    \text{true}, & \text{if }  f = h \wedge m \\
    \text{false}, & \text{otherwise}
  \end{array}
  \right.
\end{equation}
This decision operation may be used as an action to forward the packet, or for an ACL operation. A simple observation of~\eqref{eq:bf-match} and~\eqref{eq:ip4-match} shows that if the LID is used in the field and mask entries, \emph{i.e.} $f=m=L(e)$ and $h=F_p$, then the decision operation in the TCAM is identical to that required for the BF match. From a conventional viewpoint this seems unusual, as it might be expected that the wildcard mask is a contiguous set of ``1''s from the most significant bit until the length of an IP prefix. However, it is important to note that the matching operation in the TCAM actually uses an arbitrary mask. There are two reasons for this: (a) it is easier, in hardware, to perform an arbitrary mask operation with a pre-calculated prefix mask than it is to count an arbitrary number of bits from the left (this is the reason that, traditionally, masks are expressed rather than simpler CIDR prefix lengths); and, (b) ACLs are often expressed in network configurations as address/port ranges (or lists of addresses/ports) that do not naturally fit simple prefixes but can be expressed efficiently in hardware as an arbitrary mask. For an example of the latter: consider that two ``deny'' ACLs, each matching addresses \texttt{10.1.1.0/24} and \texttt{11.1.1.0/24} respectively, can be implemented as the single TCAM entry and arbitrary wildcard mask \texttt{10.1.1.1/254.255.255.0}. In many cases equipment vendors make this transparent to users by automatically merging TCAM entries where it is possible, however, increasingly vendors are making arbitrary masks available in ACLs. It should be noted that although OpenFlow (v1.2 and later) supports arbitrary masks, many controllers have not yet implemented it. For future work, the authors propose to engineer the arbitrary match and offer it to the community through the OpenDaylight project~\cite{OpenDaylight}.

The observation that an arbitrary match allows the BF switching is only one part of the solution, it is also important to understand how this is supported in switches and controllers. It should be observed that SDN switches use a variety of chipsets and thus differ in support for particular features and version of OpenFlow. Many switches utilise the Broadcom range of ASICs and the authors of this paper have confirmed that the arbitrary match is supported in these chipsets. Furthermore, arbitrary match has been a required feature of OpenFlow from v1.2. An example of a compatible set of fields that allow for an arbitrary mask are shown in~\tabref{tab:arbitary} illustrating that it is possible to support a FID of up to 384 bits. Some switches also support arbitrary match on transport layer ports offering 384+16+16=416 bits, however, this is not currently part of the OpenFlow protocol.
\begin{table}[tb]
  \centering
  \caption{Example fields supporting arbitary mask from OpenFlow v1.2}
  \vspace{0.5em}
  \begin{tabular}{|l|l|}
    \hline
    Field & Bits\\
    \hline
    Ethernet destination & 48 \\
    Ethernet source & 48 \\
    VLAN ID & 12 \\
    IPv6 source address & 128 \\
    IPv6 destination address & 128 \\
    IPv6 flow label & 20 \\
    \hline
    Total & 384 \\
    \hline
  \end{tabular}
  \label{tab:arbitary}
\end{table}

In addition to the arbitrary match, the switch also has to support sending a packet out of multiple ports to enable multicast. Hence, triggering only the action associated with the first match in a single table would not suffice, as the BF multicast switching requires the LID of each outgoing port to be tested and all the matching LIDs to have the action of \emph{output to port}. To enable this behaviour in SDN, one solution is to make use of pipelining, configured using the multiple table support feature provided by OpenFlow from v1.2. Thus, for a $d$ port switch with LIDs $L_1,\ldots,L_i,\ldots,L_d$ there are two flow entries in the $i$th flow-table as shown in~\secref{tab:tables} showing that only $2d$ flow entries are required for implementing a stateless SDN multicast solution. 
\begin{table}[tb]
  \centering
  \caption{Flow entries in the $i$th table where $i=1\ldots d$}
  \label{tab:tables}
  \vspace{0.5em}
  \begin{tabular}{|l|l|l|}
    \hline
    Priority & Match & Action\\
    \hline
    $p$ & $L_i/L_i$ & \texttt{output port}~$i$, \texttt{goto table}~$i+1$ \\
    \hline
    $p-1$ & $\texttt{any}$ &\texttt{goto table}~$i+1$ \\
    \hline
  \end{tabular}
\end{table}

However, although SDN switches (OpenFlow v1.2 or later) support tables, they support a limited number of tables, possibly only a single table. Where not enough tables are available an alternative implementation is required. One option is to construct flow entries in a single table of the form shown in~\tabref{tab:notables}, where every possible combination of outgoing LIDs is created. However, this gives rise to flow entries that grow exponentially in number with the node degree $d$; specifically the number of flow entries $\epsilon(d)$ is 
\begin{equation}
\label{eq:notable}
\epsilon(d)=\sum_{i=1}^d \binom{d}{i}=2^d-1
\end{equation}
\begin{table}[tb]
  \centering
  \caption{Flow entries in the case of one flow table}
  \vspace{0.5em}
  \begin{tabular}{|l|l|l|}
    \hline
    Priority & Match & Action\\
    \hline
    $p$ & $L_1/L_1$ & \texttt{output port} 1\\
    & $\vdots$ & \\
    $p+d$ & $L_d/L_d$ & \texttt{output port} $d$\\
    $p+d+1$ & $L_1\vee L_2/L_1\vee L_2$ & \texttt{output ports} 1,2\\
    & $\vdots$ & \\
    $p+2d-1$ & $L_1\vee L_d/L_1\vee L_d$ & \texttt{output ports} 1,$d$\\
    & $\vdots$ & \\
    $p-1+\sum_{i=1}^d \binom{d}{i}$ & $L_1\vee \ldots \vee  L_d / $ & \texttt{output ports} 1$\ldots d$\\
    &$L_1\vee \ldots \vee L_d$ & \\
    \hline
  \end{tabular}
  \label{tab:notables}
\end{table}

For small node degrees, $d$, $\epsilon(d)$ may be acceptable. However, where $d$, and thus $\epsilon(d)$ is too large, a solution is to divide the switch into a number of bridges each with a smaller number of ports, together encompassing the node degree. The bridges need to be organised such that they receive internal routing of port inputs to each bridge. This has been found to operate well in the switches tested by the authors. One observation is that, although a switch may appear to support tables, it may in fact implement the solution of~\tabref{tab:tables} using a mechanism like that shown in~\tabref{tab:notables} through TCAM manipulation in the switch software. Thus, this should be checked and, if necessary, the division into multiple bridges should be implemented. We say that a switch \emph{supports tables natively} if it does so using the method in~\tabref{tab:tables} with at least $d$ tables. It is expected that, in time, most switches will support tables natively and this implementation issue will become historical.

\subsection{Deploying  the BF multicast solution}
\label{sec:summary-bf-multicast}

The above discussion can be summarised into a deployment solution for the architecture shown in~\ref{fig:arch}:
\begin{enumerate}
\item at network deployment, or topology change, LIDs are associated with links~\cite{Jokela09}
\item through either the SDN controller, or switch management software, insert flow entries according to~\tabref{tab:tables} (or according to~\tabref{tab:notables} if tables are not supported, with switches divided into bridges as required)
\item FIDs are disseminated to NAPs using an appropriate protocol either as required, or all in advance (POINT~\cite{Trossen2015} or MPSS~\cite{Zahemszky2010} are two such mechanisms)
\item no state changes are required except to map multicast traffic into the appropriate FIDs in the NAPs.
\end{enumerate}

Two important outcomes of this mechanism are that: (a) flows are inserted proactively so that there is no delay incurred from the OpenFlow controller; and, (b) there are no state changes in the core of the network. The only state requirements are that the NAPs associate incoming multicast (or unicast flows) with an appropriate FID, but this amount of state is no different to any other multicast implementation.

Although it has been shown that BF switching can be implemented in contemporary SDN switches, the switches do not support the BF FID as a normal field. An effective solution is to \emph{overload} existing protocol header fields without changing their behaviour for existing use. As this is an intra-domain solution, this overloading could be left as a deployment decision, choosing fields from the list in~\tabref{tab:arbitary} depending on the length of BF required and existing protocols used in the network. One possible implementation that would allow full ``ships-in-the-night'' operation with existing protocols is as follows:  use the IPv6 flow, source and destination fields for the FID to give a BF of length 276 bits, use the source MAC address of the source NAP and make the destination MAC address NULL to enable discrimination of the BF switching from existing protocols. Alternatively solutions could opt for using: use the VLAN ID to discriminate; or, the VLAN ID to lengthen the BF FID; and/or the Ethernet MAC addresses could be used for all or part of the FID.

\section{Analysis of TCAM state requirements}
\label{sec:algo}

In this section, we analytically compare the performance in terms of  TCAM state size of the proposed solution with two alternative common networking switching paradigms: L2switch, as described in Section~\ref{sec:sdn-switching}; and, MPLS. MPLS implementations can use label merging to reduce the number of required labels and this is typically used in a best-effort network scenario that follows standard shortest-path routes from routing protocols; this will be defined here as MPLS-LM. An alternative MPLS solution, typically used for operator transport networks, or where QoS requires unique labels for each path, does not allow label merging; this will be defined here as MPLS-NM. The analysis used in this paper compares the \emph{unicast} state, as this can be determined for all the techniques. We consider only unicast as there is not a native multicast solution for L2switch or MPLS, rather they depend upon approaches that either: (a) use many unicast paths to ``stitch'' together multicast trees with a large amount of state; (b) send all traffic in unicast to a central node for duplication into (hopefully) shorter unicast paths; or, (c) follow a combined approach of (a) and (b) with multiple replication points~\cite{Rosen12}. Thus, determining the multicast state in the L2switch or MPLS case is not possible to define in a straightforward manner, but is an operational decision with a trade-off between routing efficiency and state; for example saving state by transmitting some (or in worst case all) traffic by unicast. This trade-off is thoroughly analysed by Martinez-Yelmo \emph{et al.} ~\cite{Martinez-Yelmo2007}. Hence, in this paper the unicast state will be determined noting that, for multicast with $g$ multicast groups: in the worst case the amount of state could be an additive increase of $g$ more than the unicast state for L2switch and MPLS, but at best (but with potentially no multicast traffic savings) it would be the unicast state. In the case of the BF implementation the multicast state is identical to the unicast state, irrespective of $g$. Note that, the state in the NAPs (mapping groups to trees) is not compared, as it is the same for any implementation, and typically of order $O(g)$.

The following will determine the expected number of TCAM entries in a switch, $\E[T]$, and a bound on the maximum number of TCAM entries in a switch, $\max(T)$. Recall that the network is represented as a graph $G(E,V)$ which has $N=|V|$ nodes. The degree of a node $v \in V$ is denoted as $d_v$, or for simplicity $d$, dropping the subscript $_v$. For the analysis it will be assumed that there is full connectivity in the network between any two NAPs where in practice a NAP will be co-located at any switching point. This is based on the observation that while not every pair of nodes will have user traffic between them, there is likely to be at least control traffic between every node (\emph{e.g.} external routing updates) and that, in practice, operators rarely implement switching except at a site where there is also a point-of-presence with ingress/egress traffic. Thus, with the set of all unicast paths $C_G=\{c_{s,t}\;|\;c_{s,t}=\langle s,\ldots, t\rangle; \; \forall s,t\in V; \; s\neq t\}$ the number of paths is $|C_G|=N(N-1)$.

Using the model defined above, the number of TCAM entries for L2switch and MPLS-NM will be the same with either a unique destination/source MAC per-path or unique MPLS label per-path respectively. To avoid duplicating analysis, the number of TCAM entries for both of these scenarios, $T_L$, will be combined.  The maximum number of TCAM entries in this case is highly topology dependent but we can upper bound it as follows: 
\begin{equation}
\label{eq:maxTL} \max(T_L) \leq N(N-1) < N^2 
\end{equation}
which corresponds to the worst case where one node carries all the traffic;  Section~\ref{sec:results} will show this bound is on average, approximately double the true value in practice.  The expected value, $\E[T_L]$, can be derived analytically from averaging across the TCAM entries of each path across the total number of nodes. Define the path length of $c_{s,t}$ as $\ell (c_{s,t})=|c_{s,t}|-1$. As each path has a unique source and destination Ethernet MAC address (or MPLS label) used to define a flow, then there is one TCAM entry per-path through each node. Thus:
\begin{equation}
  \label{eq:ETLprelim}
  \E[T_L] = \sum_{ c_{s,t}\in C_G} \frac{\ell(c_{s,t})}{N}
\end{equation}
The mean path length, $\E[\ell_G]$ is defined as:
\begin{equation}
  \label{eq:L}
  \E[\ell_G] = \sum_{ c_{s,t}\in C_G}\frac{\ell(c_{s,t})}{|C_G|}
\end{equation}
Rearranging~\eqref{eq:L} to determine $\sum_{ c_{s,t}\in C_G}
\ell(c_{s,t}) = \E[\ell_G] |C_G| = \E[\ell_G] N(N-1)$ and substituting
into~\eqref{eq:ETLprelim} gives:
\begin{equation}
  \label{eq:ETL}
    \E[T_L] =  \E[\ell_G] N(N-1)/N = \E[\ell_G] (N-1)
\end{equation}

In the case of MPLS-LM, the number of TCAM entries $T_M$ is simply determined by the observation that, with label merging, each node needs only one label going to every other node, with upstream paths being merged into this label. Thus there are the same number of labels in every node meaning that $\max(T_M) = \E[T_M] = N-1, \; \forall v \in V $.

For the proposed BF switching we note that the implementation varies depending upon whether tables are supported and, if they are not, how many bridges the switch is divided into. These implementations are denoted as $B(b)$ where $b$ is the number of bridges on a switch with $d$ ports, giving a range: from $B(1)$ to $B(d)$.  We define $B(1)$ as one bridge with no multiple table support; while we define $B(d)$ as either one bridge per-port, or a switch supporting multiple tables natively. For BF switching, using an implementation as proposed in Section~\ref{sec:summary-bf-multicast}, switches generally require two TCAM entries per-flow, compared to a single TCAM entry per-flow for L2switch or MPLS, due to the wider matching field; consequently the analysis will include an additional factor of two for each flow entry. To take into account spreading the ports, as equally as possible, over $x$ bridges the analysis of~\eqref{eq:notable} gives the number of TCAM entries for BF switching, $T_{B(x)}$, in a node of degree $d$ as:
\begin{align}
T_{B(x)}= &2(\bmod(d,x)) (2^{\lceil d/x \rceil}-1) \nonumber \\
 &+ 2(x- \bmod(d,x)) (2^{\lfloor d/x \rfloor}-1)\nonumber \\
= & (x+\bmod(d,x)) 2^{\lfloor d/x \rfloor +1} -2x
\label{eq:Tb}
\end{align}
where $\mod(d,x)$ is $d$ expressed as modulus $x$. For most physical (router level) topologies (including those used in the results later) it
has been shown that $d$ typically obeys a discrete Weibull
distribution with shape parameter of 0.42~\cite{Spring02}. Unfortunately, in this case, the analytical solution of the expected value  $\E[T_{B(x)}]$, in the general case, does not converge, as the probability density function of $d$ is an exponent raised to power $d$ and the Weibull distribution exponent is of power $-d^k$; thus $\E[T_{B(x)}]$ would only converge if $k>1$. This value of $k$ is much higher than observed in real networks.
The maximum value in the BF case, $\max(T_{B(x)})$, is determined by the maximum node degree in the network $\max(d_v)$:
\begin{equation}
\max(T_{B(x)}) \leq (x+\bmod(\max(d),x)) 2^{\lfloor \max(d)/x \rfloor +1} -2x
\label{eq:maxTb}
\end{equation}
Notably,  in the case that $B(d)$, \ie one port per-bridge, or OpenFlow tables natively implemented, then: as $T_{B(d)}=2d$ it follows that $E[T_{B(d)}]=2E[d]$ and $\max(T_{B(d)})=2\max(d)$.

\section{TCAM state required in realistic networks}
\label{sec:results}

The TCAM state required in a range of networks was modelled using the assumption of full connectivity and counting the required TCAM entries for the assigned paths. The modelling was carried out using both synthetic network topologies and real topologies published by the Internet Topology Zoo (ITZ)~\cite{Knight11}. The synthetic network topologies were created with a node degree following a Weibull distribution with shape parameter 0.42, according to a survey of ISP router level topologies~\cite{Spring02}. The node degree of the ITZ was found to fit a Weibull distribution with shape parameter 0.44 (using non-linear least squares with exponential weights) and thus, statistically speaking, the synthetic networks resemble the ITZ networks. As with the analysis in Section~\ref{sec:algo}, the TCAM counts are for unicast paths.  For BF switching the values are the same as the unicast values, irrespective of the number of multicast groups. For the others, the state could additively increase in each switch by up to $g$, the number of multicast groups. The synthetic networks are implemented with $N=20,40,\ldots,200$ and with 10 repeats. For clarity in the plots, the results for the 225 ITZ networks are aggregated to the nearest multiple of ten. The exception to this is the single, largest, ``Kdl'' network, with $N=754$, which is shown separately at the end.

\begin{figure}[tb]
  \centering
\includegraphics[width=\linewidth]{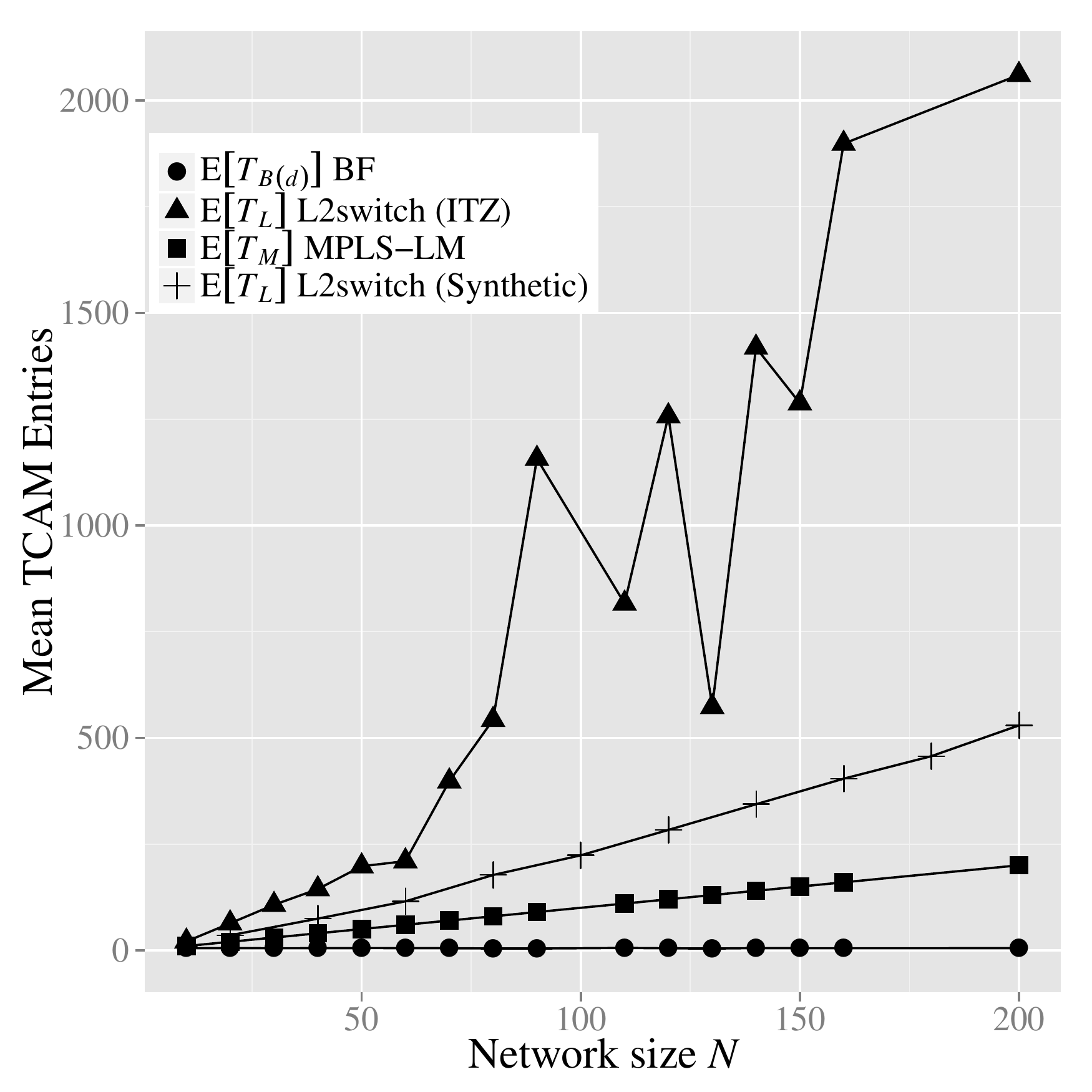}
  \caption{Mean TCAM entries using synthetic and ITZ networks. For
    each of MPLS-LM and BF $B(d)$ only one line is shown as the difference between
    synthetic and ITZ is not visible.}
  \label{fig:synthMean}
\end{figure}

The mean TCAM entries are shown in~\figref{fig:synthMean}, showing the values for both synthetic and ITZ topologies. The initial sets of results show the BF implementation using either OpenFlow tables or with one bridge-per-port ($B(d)$). The results for the MPLS-LM and BF switching are so close for the both ITZ and synthetic results that only one line is plotted. It is notable that the L2switch (also MPLS-NM) results are significantly higher than both the MPLS-LM and BF switching. Mean values for BF are too small to see on the plot, they vary between 3.6 and 5.5 TCAM entries. The L2switch values in the ITZ network are generally much higher than for the synthetic networks as the mean path length is generally larger in the ITZ networks. This demonstrates that node degree itself is not enough to model realistic networks. However, the synthetic results show that the trend is proportional to node degree as predicted in Section~\ref{sec:algo}. For the remainder of the results only the ITZ results are shown.

\begin{figure}[tb]
  \centering
\includegraphics[width=\linewidth]{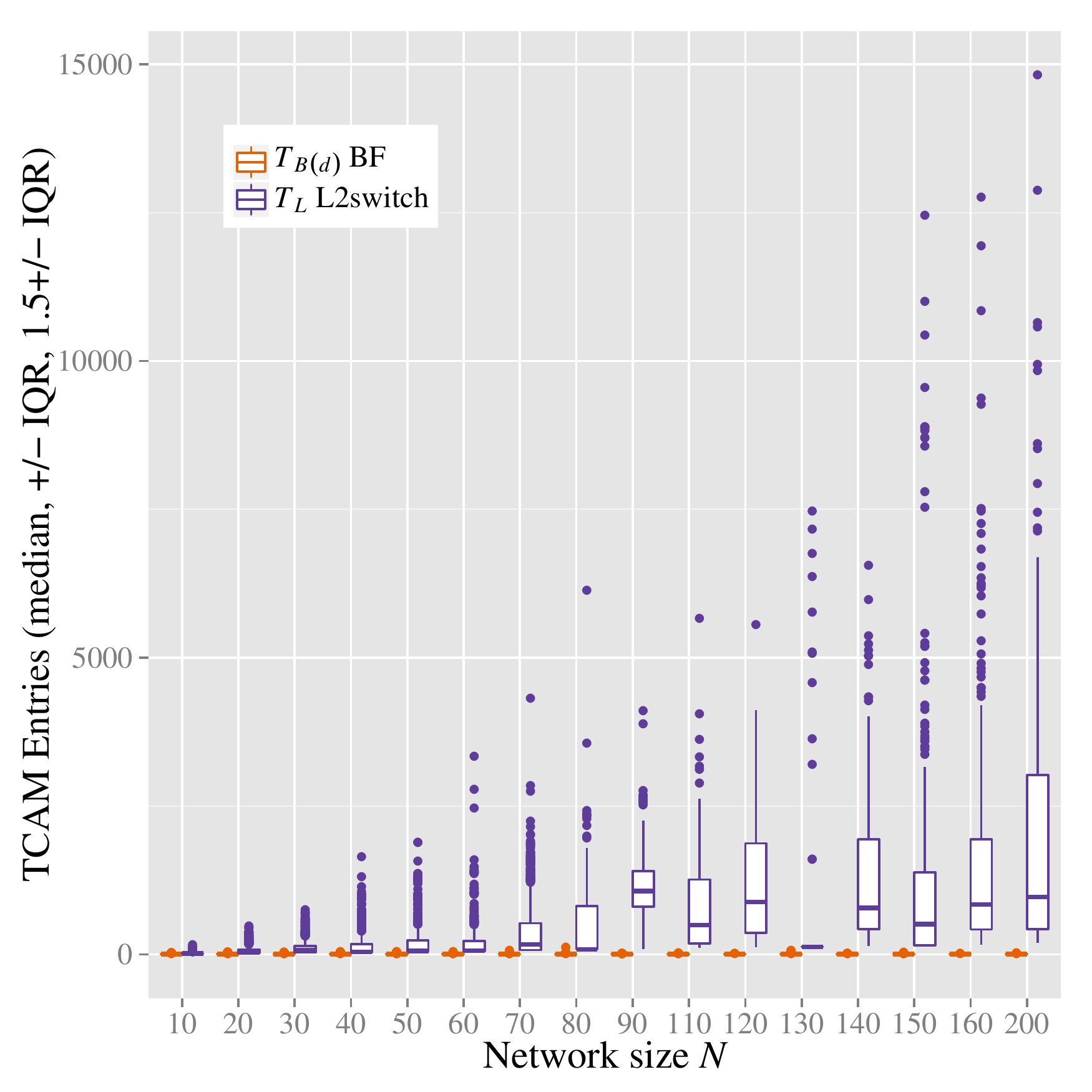}
  \caption{Comparing the number of  TCAM entries for BF $B(d)$ and L2switch
    in ITZ networks}
  \label{fig:zooBoxPlotL2}
\end{figure}
\begin{figure}[tb]
  \centering
\includegraphics[width=\linewidth]{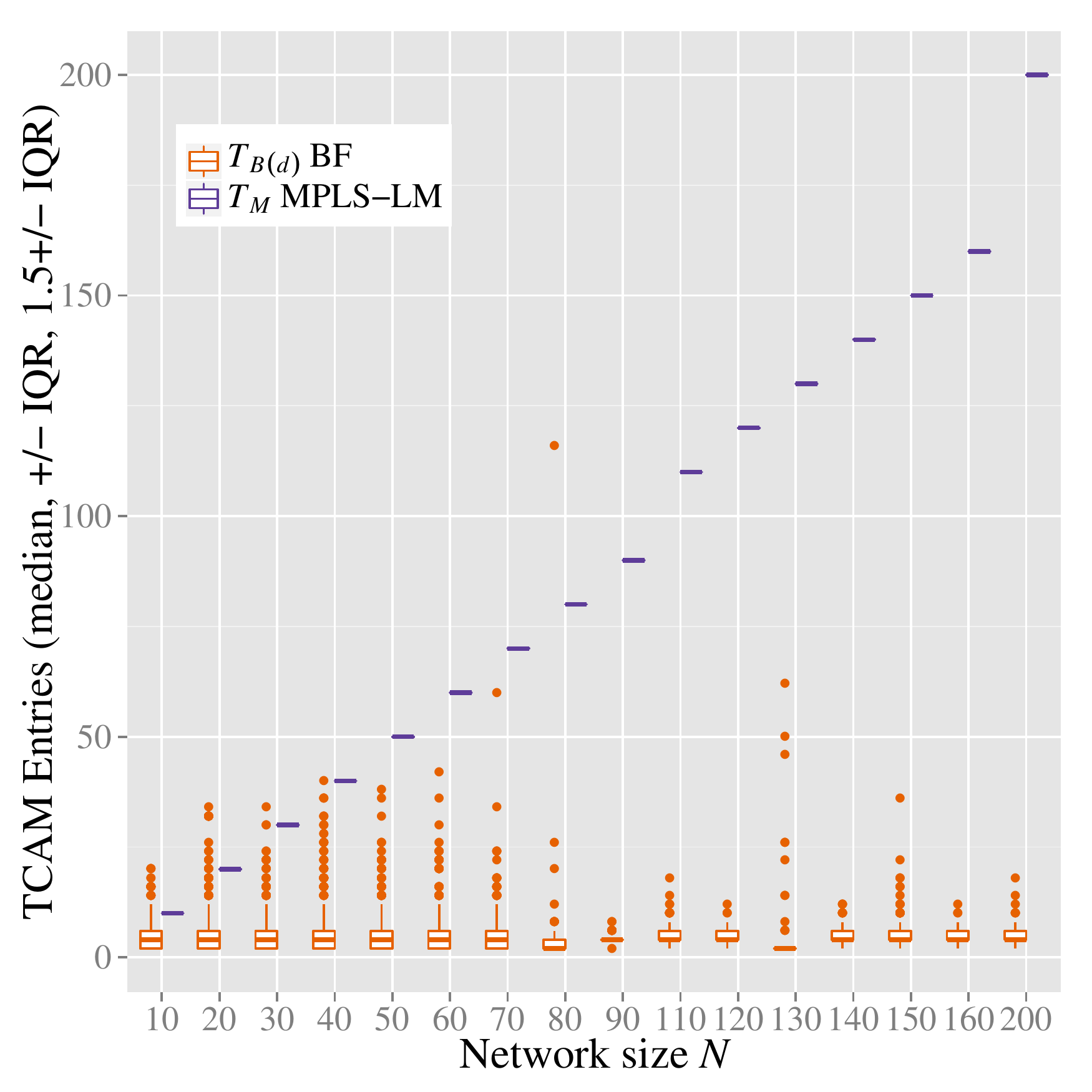}
  \caption{Comparing the number of  TCAM entries for BF B(d) and MPLS-LM
    in ITZ networks}
  \label{fig:zooBoxPlotMPLS}
\end{figure}
The distribution of the TCAM entries is heavily skewed with some values considerably higher than the mean, as shown in~\figref{fig:zooBoxPlotL2} for the L2switch and BF results; This demonstrates that in some cases, the L2switch (MPLS-NM) results are very large in some switches. Generally the results for L2switch show that prediction of the highest TCAM entries $\max(T_L)$ from~\eqref{eq:maxTL} was conservative with the average value $\E[\max(T_L)]=0.54 N(N-1)$, but in one case it was as high as $0.92 N(N-1)$.  \figref{fig:zooBoxPlotMPLS} shows the comparison between the BF and MPLS-LM scenarios showing that, in general, the TCAM requirements for BF are considerably smaller with only a few switches requiring more TCAM entries. This figure shows that there is one outlier in the BF case, corresponding to the  ``Ulaknet'' network with $N=82$, which has one node requiring significantly more TCAM entries than any other. It is useful to consider this case as it demonstrates the influence of node degree on the number of TCAM entries for BF switching. The outlying node has a degree of $d=58$, this is highly unusual in a physical level topology, although not in a logical topology such as BGP peering, which is not relevant here. In this case BF values are $\max(T_{B(d)})=116$ when the switch supports tables (or is divided into $d$ bridges) and $\max(T_{B(8)})=2544$ when the switch is divided into 8 bridges; this can be compared to the L2switch/MPLS-NM value of  $\max(T_L=6133)$. This is an example where it is necessary, in the BF case, to use a switch that either has native support for OpenFlow tables, or that can be subdivided into a suitable number of bridges. In the case where neither of these would be possible there would be of the order $2^{58}$ entries required, not remotely feasible. In this set of results, although MPLS-LM has a larger number TCAM entries than BF, for the majority of cases, it does at least have the advantage of consistent and predictable numbers as it is simply equal to the number of nodes in the network.

\begin{figure}[tb]
  \centering
\includegraphics[width=\linewidth]{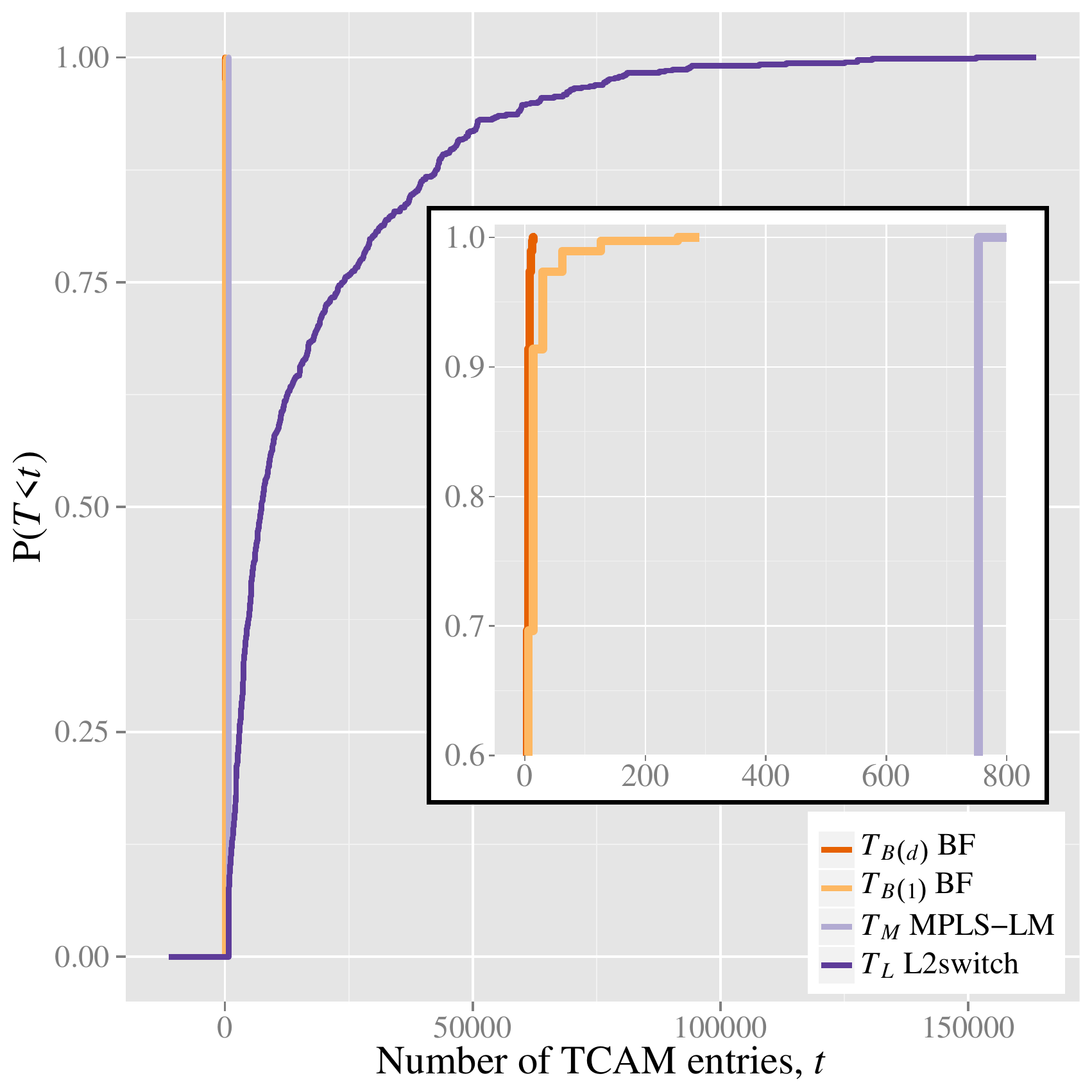}
\caption{Empirical cdf of TCAM entries for the ITZ Network
  ``Kdl''.}
  \label{fig:cdfKdlL2}
\end{figure}

The cumulative distribution function (cdf) for the largest ITZ network, ``Kdl'' is shown in~\figref{fig:cdfKdlL2}, which demonstrates that a large number of nodes need a very large number of TCAM entries for the L2switch/MPLS-NM cases. However, the BF switching needs very few entries; indeed even without table support, or splitting any of the switches into bridges, the largest number of TCAM entries is 254 for one node and with 95\% of the nodes needing 30 or less. With table support the maximum for BF switching is 14 entries and 90\% of the nodes using 6 or less TCAM entries. This leaves plenty of TCAM entries for other existing protocols, or paves the way for switches with much reduced TCAM requirements and thus increased power savings.

\section{Conclusion and Future Work}
\label{sec:conclusion}
A stateless multicast solution for SDN has been presented. The proposal builds on previous efforts to use Bloom Filters for multicast switching but shows, for the first time, that it can be deployed in contemporary SDN switches without any changes. The state in the TCAM of the switches in realistic network topologies are analysed showing that this multicast mechanism has significantly less state than unicast switching of the most common alternatives such as layer-2 switching or MPLS. Furthermore, the mechanism only requires SDN flow entries to be inserted proactively, significantly solving the scalability and delay considerations of solutions  that require reactive insertion from the centralised SDN controller.

While the proposed solution has been shown to work with existing SDN switches, there are two areas where the work should be extended. The first is to extend suitable SDN controllers to support the arbitrary match that the OpenFlow protocol requires and the authors intend to do this through the OpenDaylight project. A second extension of the mechanism is the resolution of the scalability problem caused by the false positives inherent to Bloom Filters. One suggested proposal is to divide the network into zones of suitable sizes so that a larger network can be implemented using a subset of smaller zones that have no false positives. Future work will model this solution to demonstrate that this stateless multicast solution for SDN can scale to any network size.

\bibliographystyle{IEEEtran}
\bibliography{refs}

\end{document}